# Cavity-enhanced optical frequency comb spectroscopy in the mid-infrared – application to trace detection of $H_2O_2$

Aleksandra Foltynowicz[*], Piotr Masłowski[a], Adam J. Fleisher, Bryce Bjork, and Jun Ye

*JILA, National Institute of Standards and Technology and University of Colorado,*

*Department of Physics, University of Colorado, Boulder, Colorado 80303-0440, USA*

*fax: + 1 (303) 492-5235,*

*email: aleksandra.matyba@jila.colorado.edu*

We demonstrate the first cavity-enhanced optical frequency comb spectroscopy in the mid-infrared wavelength region and report the sensitive real-time trace detection of hydrogen peroxide in the presence of a large amount of water. The experimental apparatus is based on a mid-infrared optical parametric oscillator synchronously pumped by a high power Yb:fiber laser, a high finesse broadband cavity, and a fast-scanning Fourier transform spectrometer with autobalancing detection. The comb spectrum with a bandwidth of 200 nm centered around 3.75 µm is simultaneously coupled to the cavity and both degrees of freedom of the comb, i.e., the repetition rate and carrier envelope offset frequency, are locked to the cavity to ensure stable transmission. The autobalancing detection scheme reduces the intensity noise by a factor of 300, and a sensitivity of $5.4 \times 10^{-9}$ cm$^{-1}$ Hz$^{-1/2}$ with a resolution of 800 MHz is achieved (corresponding to $6.9 \times 10^{-11}$ cm$^{-1}$ Hz$^{-1/2}$ per spectral element for 6000 resolved elements). This yields a noise equivalent detection limit for hydrogen peroxide of 8 parts-per-billion (ppb); in the presence of 2.8% of water the detection limit is 130 ppb. Spectra of acetylene, methane and nitrous oxide at atmospheric pressure are also presented, and a line shape model is developed to simulate the experimental data.

---

[a] Current address: Instytut Fizyki, Uniwersytet Mikołaja Kopernika, Toruń, Poland

# 1. Introduction

Laser absorption spectroscopy is an important method used for research in atomic and molecular physics and frequency metrology. It has been successfully applied to issues of fundamental scientific importance and serves as the backbone for numerous applications in physics, chemistry, and biology. These applications include control of manufacturing processes [1-3], detection of poisonous gasses or explosives [4, 5], human breath analysis for medical applications [6-9], environmental monitoring [10-13], and optical methods of $^{14}$C dating [14]. Laser spectroscopy also provides high resolution and precision laboratory data essential for satellite remote sensing of planetary atmospheres and for monitoring of climate changes in Earth's atmosphere [15]. Pioneering advances in laser light sources, measurement techniques, and detection systems have opened new windows of opportunity for many of those applications mentioned above. Measurement noise has been reduced and absorption sensitivity increased via modulation techniques [16, 17], the use of high finesse cavities [18], and the combination of the two in the NICE-OHMS method [19, 20]. These efforts make possible impressive shot-noise-limited absorption sensitivities [19], very high signal-to-noise ratios [21] and the capability to measure ultra-weak molecular transitions [22].

Many applications, such as atmospheric chemical dynamics, pollution detection and human breath analysis, depend crucially on the ability to monitor absorption of different molecular species with high spectral resolution and in a short acquisition time. Direct absorption spectroscopy based on continuous wave (cw) laser sources can offer high sensitivity, however their intrinsically narrow bandwidth necessitates a scanning time proportional to the desired total measurement range. Alternatively, techniques based on broadband sources of incoherent radiation like traditional Fourier transform infrared spectroscopy allow measurements in a wide spectral range. However, their spectral resolution is limited by the monochromator or maximum optical path difference in a Fourier transform spectrometer [23]. Due to the low spatial coherence of the light sources this method also requires long averaging times in order to reach sensitivities comparable to those offered by cw-laser based systems.

Optical frequency comb (OFC) sources eliminate these limitations by combining broad spectral bandwidth with high spectral resolution. An optical frequency comb results from interference between the periodic femtosecond pulses of the coherent pulse train emitted by a mode-locked laser. The broadband (up to an octave) spectrum of the laser in the frequency domain consists of thousands of narrow, equidistant, discreet lines, or comb modes, separated by the laser repetition rate ($f_{rep}$). The frequencies of single comb lines are shifted from integer multiples of $f_{rep}$ by the carrier envelope offset frequency ($f_{ceo}$), caused by different average group and phase velocities in the laser cavity. As a result, the frequency of the *nth* comb mode is given by $v_n = nf_{rep} + f_{ceo}$, where *n* usually is in the order of $10^5$-$10^6$. By controlling the radio frequencies $f_{rep}$ and $f_{ceo}$ it is possible to precisely control the frequency of each comb line [24]. This property is utilized when optical frequency combs act as precise frequency rulers for measurements of the frequency of a cw laser [25-28].

The recently developed technique of direct frequency comb spectroscopy (DFCS) [29] directly employs optical frequency combs to probe spectral features in a massively parallel fashion [30]. It offers the entire optical bandwidth of the femtosecond laser combined with a resolution limited by the width of a single comb line (typical values range from several kHz to sub-Hz [29, 31, 32]). Providing high spectral brightness and spatial coherence, the DFCS technique is virtually equivalent to simultaneous measurements with tens of thousands of frequency-stabilized narrow-linewidth lasers. Once comb teeth are resolved, there is no trade-off between the time of the measurement and spectral resolution. The sensitivity of the DFCS technique can be greatly enhanced through the use of a high finesse optical cavity, which increases the interaction length with the sample [30, 33]. Due to the similarity of the cavity mode structure to the OFC spectrum, the femtosecond laser light can be efficiently coupled into the optical cavity by proper control of the two radio frequencies $f_{rep}$ and $f_{ceo}$ [24]. Several detection schemes for cavity-enhanced (CE-) DFCS have been implemented [33-35], based either on one- [36, 37] or two-dimensional dispersion elements [34], on the Vernier method [38], or on Fourier transform spectroscopy [39, 40]. When dispersion elements are used to resolve the spectrum transmitted through the cavity, the frequency-to-amplitude noise conversion by the narrow cavity modes can be circumvented by measurement of cavity ringdown [37] or by fast modulation of either the comb modes or the cavity resonance frequencies while detecting the averaged cavity output [30, 34, 36]. This approach cannot be implemented in combination with Fourier transform spectroscopy (FTS), which requires constant transmission through the cavity. The OFC-based FTS spectrometers are based either on a Michelson interferometer with a mechanical translation stage [35, 40-42] or on dual comb spectroscopy [39, 43-45], which utilizes two femtosecond lasers with slightly different repetition rates, thus mimicking the effect of a fast-scanning delay stage. In this work we implement a recently developed detection scheme that efficiently suppresses the amplitude noise by combining an OFC tightly locked to a high finesse cavity and a fast-scanning FTS with autobalancing detector, allowing hours of uninterrupted operation [40].

All previous realizations of CE-DFCS have been in the visible or near-infrared wavelength range, where commercial OFC sources are now available. Progress towards longer wavelengths has recently yielded new direct OFC sources, including the Th:fiber laser operating at 2 µm [46] and the Cr2+:ZnSe laser operating at 2.4 µm [47]. Longer wavelengths can also be reached via nonlinear processes, either by Raman shifting [48], difference frequency generation [49-51] or with optical parametric oscillators [52-55]. The latter two provide spectral access to the important molecular fingerprint region located above 3 µm.

DFC spectrometers, both employing multi-pass cells and optical enhancement cavities, have already been used for detection of numerous molecular species important for environmental research [37, 42, 56, 57], monitoring of production processes [48, 58], and have enormous potential as a tool for non-invasive human breath analysis [34, 42]. In the latter application, the ability to measure accurate broadband absorption spectra in real time will allow discriminating between more than 500 molecular species present at different concentrations in human breath [7]. Very often an absorption spectrum of the molecule of interest is overlapped with a spectrum of a more abundant or stronger absorbing species. One example is the hydrogen peroxide ($H_2O_2$) molecule, whose increased presence in human breath could indicate oxidative stress in the lungs [59, 60]. Detection of elevated levels of $H_2O_2$ can be used for early diagnosis and monitoring of such diseases as asthma [61-63], chronic obstructive pulmonary disease [64], or acute respiratory distress syndrome [65, 66], a severe inflammatory condition with up to 50% mortality [67]. The detection of $H_2O_2$ in human breath is most commonly performed by the creation of breath condensate and the use of spectrophotometric or spectrofluorometric methods, which achieve detection limits in the range of a few ppb to a few hundred ppb [60]. A commercially available biosensor offers detection limits on the order of a few tens of ppb, but requires everyday calibration [68]. The need for breath condensation and the use of chemical reactions for analysis precludes the use of these methods for real-time applications. High resolution laser spectroscopy has recently been recognized as a tool for quantitative breath analysis in gas phase, eliminating the need of sample condensation and pre-filtering [8, 9]. However, up to now no laser-spectroscopy-based system for optical detection of $H_2O_2$ in human breath has been reported despite the molecule's importance as a potential marker for numerous diseases.

We report here a mid-infrared CE-DFCS system dedicated to the detection of hydrogen peroxide in human breath, based on an optical parametric oscillator (OPO), a high finesse low dispersion enhancement cavity and a fast-scanning Fourier transform spectrometer with autobalancing detection. The OPO comb is tightly locked to the cavity, and we describe in detail the method of two-point locking employed to achieve wide optical bandwidth and high stability in transmission through the cavity. We discuss the absorption line shapes and evaluate the performance of the system by measurements of acetylene, methane and nitrous oxide spectra. We have chosen to detect $H_2O_2$ at the intercombination band $\nu_2 + \nu_6$ at 3.7 µm, which is far detuned from strong water absorption bands [69], in order to enable detection of $H_2O_2$ in the presence of water vapor at a percentage level. We obtain detection limits superior to other $H_2O_2$ detection methods used for breath analysis, as the time needed for sample preparation and data acquisition and analysis in our approach is three orders of magnitude shorter.

## 2. Absorption line shapes

The electric field of a single comb line transmitted through a cavity containing an absorbing sample, $E_t(\nu)$, is given by [70]

$$E_t(\nu) = E_{inc}(\nu) \frac{t(\nu) e^{-i\varphi(\nu)/2 - \delta(\nu)L - i\phi(\nu)L}}{1 - r(\nu) e^{-i\varphi(\nu) - 2\delta(\nu)L - 2i\phi(\nu)L}}, \tag{1}$$

where $E_{inc}(\nu)$ is the incident electric field, $t(\nu)$ and $r(\nu)$ are the frequency-dependent intensity transmission and reflection coefficients of the cavity mirrors, respectively, $L$ is the cavity length [cm], $\varphi(\nu)$ is the round trip phase shift in the cavity, given by $4\pi\nu nL/c$, where $n$ is the refractive index of the buffer gas and $c$ is the speed of light; and where $\delta(\nu)$ and $\phi(\nu)$ are the attenuation and phase shift of the electric field due to the analyte per unit length, given by

$$\delta(\nu) = \frac{Sn_A}{2} \operatorname{Re} \chi(\nu), \tag{2}$$

$$\phi(\nu) = \frac{Sn_A}{2} \operatorname{Im} \chi(\nu). \tag{3}$$

Here, $S$ is the molecular linestrength [cm$^{-1}$/(molecule/cm$^{-2}$)], $n_A$ the density of absorbers [molecules/cm$^3$], and $\chi(\nu)$ is the complex line shape function [cm], in the pressure broadened regime given by the Voigt profile (i.e., error function of a complex argument). The frequency dependent transmitted intensity $I_t(\nu)$, calculated as $E_t(\nu) E_t^*(\nu)$ and normalized to the intensity in the absence of the analyte, $I_0(\nu)$, is therefore given by [40]

$$\frac{I_t(\nu)}{I_0(\nu)} = \frac{t^2(\nu)e^{-2\delta(\nu)L}}{1-r^2(\nu)e^{-4\delta(\nu)L}-2r(\nu)e^{-2\delta(\nu)L}\cos\left[2\phi(\nu)L+\varphi(\nu)\right]}. \qquad (4)$$

In cw locked cavity-enhanced absorption spectroscopy, the laser frequency is stabilized to the center of a cavity mode, so that the round trip intracavity phase shift is equal to $2q\pi$, where $q$ is an integer mode number. If the field attenuation due to the analyte is smaller than the intracavity losses, Eq. (4) can be series expanded yielding the familiar expression

$$\frac{I_t(\nu)-I_0(\nu)}{I_0(\nu)} = \frac{2F(\nu)}{\pi}\alpha(\nu)L, \qquad (5)$$

where $\alpha(\nu) = 2\delta(\nu)$ is the single pass absorption per unit length, equal to twice the attenuation of the electric field, and $F(\nu)$ is the cavity finesse, defined as $\pi\sqrt{r(\nu)}/[1-r(\nu)]$.

When a frequency comb is coupled into a cavity, achieving a good match of the comb and cavity spectra is usually possible only over a limited wavelength range, determined by the dispersion of the cavity mirrors, which causes the cavity FSR to also vary with optical frequency. As a result, some comb lines are not in exact resonance with their respective cavity modes and instead line up at the slopes of the cavity modes. For these comb lines the round trip intracavity phase shift is equal to $\varphi(\Delta\nu) = 2q\pi + 2\pi\Delta\nu/FSR$, where $\Delta\nu$ is the frequency detuning of the comb line from the center of the cavity mode, and $FSR = c/(2Ln)$ is the cavity free spectral range.

Absorption line shapes calculated using Eq. (4) for comb lines locked to the centers of cavity resonances as well as to the slopes are shown in Fig. 1. The cavity and molecular line parameters used for the calculations are: cavity finesse of 3800 and FSR of 273 MHz (i.e., cavity linewidth of 36 kHz), molecular linestrength of $10^{-20}$ cm$^{-1}$/(molecule cm$^{-2}$), Doppler width of 96 MHz, pressure broadening of 3 MHz/torr, intracavity pressure of 760 torr and concentration of the absorbing molecule of 5 ppm. The black curve shows the on-resonance case, while the red dashed and blue dash-dotted curves are calculated for the cases of a detuning $\Delta\nu$ of 10 and 20 kHz (i.e. 1/3 and 2/3 of the cavity linewidth), respectively. For a nonzero detuning the power transmitted through the cavity (i.e., the baseline) decreases, the absorption line shapes become asymmetric and the relative absorption depth decreases.

## 3. Experimental setup and procedures

The system, schematically depicted in Fig. 2(a), is based on a femtosecond fiber-laser-pumped mid-infrared optical parametric oscillator (OPO). The comb output is locked to an external high finesse cavity using a two-point stabilization scheme based on the Pound-Drever-Hall (PDH) method [71]. Phase modulation for the PDH error signal is obtained by dithering the pump laser cavity length using a fast PZT at one of its resonance frequencies (760 kHz). The light reflected from the cavity is picked off with a beamsplitter (BS), dispersed with a reflection grating and directed on two photodiodes (PD1 and PD2). Each photodiode detects approximately 1 nm of light resonant with the cavity around two locking points separated by tens of nm. The signals from the two photodiodes are demodulated at the dither frequency to yield error signals fed into respective servo systems.

The light transmitted through the cavity is coupled into a home-built fast-scanning Fourier transform spectrometer (FTS) [35, 40, 42], shown in detail in Fig. 2(b), equipped with an autobalancing HgCdZnTe (MCZT) detector (PD4). The interferogram measured at the balanced output of the detector is digitized with a 22 bit data acquisition board (DAQ) at a rate of 3 Msample/s. Frequency scale is calibrated with the use of an ultrastable nonplanar ring oscillator (NPRO) at 1064.457 nm (with long term frequency stability of $10^{-7}$), whose beam is propagating parallel to the mid-IR beam through the FTS. The interferogram of the reference laser is recorded by an InGaAs photodiode (PD3) and high-pass filtered, and the zero crossings of this interferogram are used as accurate optical path difference markers, at which the mid-IR interferogram is resampled.

### 3.1. Mid-infrared OPO

The mid-IR non-degenerate OPO [53] is synchronously pumped by a high-power Yb:fiber femtosecond laser [72] with a repetition rate of 136.6 MHz. The center wavelength of the idler can be tuned between 2.8 and 4.8 µm and the simultaneous spectral coverage ranges from 100 to 300 nm depending on the center wavelength. The center wavelength tuning is achieved by vertically translating a periodically poled $LiNbO_3$ crystal with a fan-out structure within the pump beam path. At 3.7 µm, the simultaneous bandwidth is 150 nm and the output power is set to 400 mW.

The comb properties of the OPO have been demonstrated in Ref. [53], where the two degrees of freedom of the OPO comb have been stabilized to optical and radio frequency references. In the process of synchronous pumping, the repetition rate of the OPO is equal to the repetition rate of the pump laser. The $f_{rep}$ of the pump laser can be changed by tuning the fiber laser cavity length. Slow but large range tuning is done with a fiber stretcher and fast modulation can be applied via a PZT with a bandwidth of 100 kHz, on which the semiconductor saturable absorber mirror (SESAM) is mounted. Changing the pump laser $f_{rep}$ causes a change in the OPO idler $f_{ceo}$, due to dispersion inside the OPO cavity [73]. The $f_{ceo}$ of the OPO idler can also be controlled by changing the OPO cavity length [53, 74] via two PZTs on which the OPO cavity mirrors are mounted, one slow with a large range, and the other fast with a bandwidth of 100 kHz [75]. Below we demonstrate for the first time how the femtosecond OPO can be stabilized to an external enhancement cavity by controlling both degrees of freedom of the comb.

### 3.2. High finesse broadband cavity

The length of the enhancement cavity is 54.9 cm, i.e., half of the OPO cavity length. This yields a cavity FSR of 273.2 MHz, equal to twice the repetition rate of the OPO, which implies that every second comb tooth is coupled into the external cavity. Cavity mirrors for the sensitive detection of $H_2O_2$ were designed to have highest reflectivity and zero dispersion at 3.7 µm, where the strongest line in the targeted absorption band of $H_2O_2$ is located. The cavity finesse as a function of wavelength was determined from a cavity ringdown measurement. A fast sweep of the comb lines across cavity resonances was achieved by modulating the PZT in the fiber pump laser. A monochromator was placed in the beam path behind the cavity to enable measurements with spectral resolution of 0.2 nm. Figure 3 shows the measured cavity finesse (blue circular markers) together with a fitted 3$^{rd}$ order polynomial (red curve). The peak finesse, obtained from the fit, is 3800 at 3815 nm, and the finesse is relatively constant (decreasing by less than 3%) in the range between 3730 and 3900 nm. This indicates that the FSR is relatively constant in this wavelength range as well [76].

The back surface of the mirrors is wedged (1 minute) and antireflection coated to minimize multiple reflections inside the mirror. The mirror losses around the center wavelength were determined from a measurement of power transmitted and reflected on cavity resonance when the OPO was locked to the cavity. In general, only 36.5% of the power incident on the cavity (200 mW) is coupled into the $TEM_{00}$ mode of the cavity. First of all, half of the comb power is rejected by the cavity due to the cavity FSR being equal to twice $f_{rep}$. Additionally, 9% of the incident power is in the modulation sidebands used for the PDH locking, and 18% of power is coupled to transverse cavity modes because of the ellipticity of the OPO beam. The on-resonance cavity transmission and reflection were measured to be 6% and 57% of the power coupled into the $TEM_{00}$ mode, respectively. From these measurements the losses of the mirrors were determined to be 630 ppm.

The mirrors are mounted in aluminum holders and glued to a glass spacer tube with a 0.8" diameter. A low-voltage ring PZT is glued between the tube and the output mirror to allow tuning of the cavity length. The length of the spacer is such that the cavity can be operated at near atmospheric pressure in the temperature range between 20 and 40 ºC. Glass was chosen as the spacer material to minimize undesired reactions with $H_2O_2$ that could be caused by a metal surface. The spacer tube is connected to a 1/4" gas flow system and a scroll pump for the delivery of methane, acetylene and nitrous oxide into the cavity. Alternatively, a glass container with solution containing $H_2O_2$ can be connected to the spacer tube from below, bringing the liquid level within 6 cm of the laser beam (the distance defined by the length of the connectors between the spacer and the container). For the $H_2O_2$ measurements the container and the cavity spacer tube are heated to 37 ºC, and the mirrors are held at a temperature of 40 ºC to prevent vapor condensation.

### *3.3. Two-point locking*

In order to ensure efficient coupling of the comb light to the cavity the two degrees of freedom of the comb must be controlled: $f_{rep}$ has to be matched to the cavity FSR, and $f_{ceo}$ has to be adjusted so that the comb lines are on resonance with the cavity modes [30, 33]. One solution would be to stabilize the comb to RF or optical references, and then adjust the external cavity length so that the resonance condition is achieved. The disadvantage of this approach lies in the potential difficulty of maintaining the correct $f_{ceo}$ for different values of cavity FSR (e.g. when the pressure inside the cavity is changed). In such configuration the idler $f_{ceo}$ would be measured via a beat note between the pump supercontinuum and the sum signal of the pump and idler [53, 74]. Measuring the beat note requires band-pass filtering around the particular frequency at which it appears, and as the correct value of $f_{ceo}$ changes, the filter would have to be tuned, which is not practical. Instead, one can slave the comb to the cavity directly, so that $f_{rep}$ and $f_{ceo}$ are automatically adjusted to the correct values. A stabilization scheme based on error signals derived at multiple wavelengths was developed by Jones *et al.* for frequency standards applications [77]. A similar two-point locking scheme was recently developed for spectroscopic applications and used for locking of an Er:fiber comb to a high finesse cavity [40]. The latter scheme is implemented here, in what is the first demonstration of a mid-infrared OPO-based optical frequency comb stabilized to an external enhancement cavity.

The mid-IR comb is locked to the external cavity using error signals derived at two different wavelengths. The error signal from PD2 (at a shorter wavelength) is used to feed back to the pump laser cavity length; it is integrated and sent to the fast PZT in the fiber laser, and the correction signal is further slowly integrated and sent to the fiber stretcher. The second error signal, obtained from PD1 at a longer wavelength, is used to feed back to the OPO cavity length. It is integrated and fed to the fast PZT, and the correction signal is further slowly integrated and sent to the slow PZT in the OPO cavity. The two-point lock is achieved by first enabling the pump laser lock (which affects both $f_{rep}$ and $f_{ceo}$), then manually tuning the external cavity length (or alternatively the OPO cavity length) so that the comb lines at longer wavelengths come into resonance with the cavity modes and finally enabling the OPO lock, which stabilizes $f_{ceo}$ to the correct value.

For measurements of $H_2O_2$ (as well as $CH_4$ and $C_2H_2$), the OPO spectrum is tuned to the center wavelength of 3760 nm, where the simultaneous OPO bandwidth is 150 nm. The blue dashed curves in Fig. 4(a) show the spectrum in cavity transmission when only the pump laser lock is enabled and the wavelength of the locking point is successively tuned across the OPO spectrum. The tuning is easily achieved without breaking the lock by turning a mirror in front of the PD2 detector (alternatively by translating the detector) so that different parts of the dispersed spectrum are incident on the detector. For these spectra, $f_{ceo}$ and $f_{rep}$ were chosen so that only a narrow portion of the spectrum (the comb lines nearest to the locking point) was transmitted through the cavity. The red dotted and black dash-dotted curves show spectra at the locking points that were chosen for pump laser lock (at 3700 nm) and for the OPO lock (at 3820 nm) and the solid blue curve shows the cavity transmitted spectrum when both locks are enabled. In general, the locking points should be separated as far apart as possible to provide better stability, but within the range in which the cavity FSR is relatively constant, so that comb lines within the two wavelengths can simultaneously be on resonance with their respective cavity modes. Comparing the solid blue curve with the maxima of the dashed blue curves reveals that at some wavelengths (e.g. below 3680 nm) the power transmitted through the cavity with the two-point lock is lower than with only one lock enabled. This is caused by the dispersion in the mirrors, which changes the FSR and makes the cavity modes walk off the comb modes, so that these comb modes line up at slopes of their respective cavity modes. The detuning of the comb lines from the cavity resonances as a function of wavelength is shown in the upper panel of Fig. 4(a) (see discussion below).

It should be noted that the curve showing cavity transmission with the two-point lock (solid blue) has to be multiplied by 1.6 in order to overlap with the maxima of single-point lock spectra (dashed blue). The decrease of power when both locks are enabled is caused by larger amplitude noise, originating from an increased frequency-to-amplitude noise conversion when more comb lines are resonant with the cavity modes. In the case of a single-point lock, the comb lines adjacent to those that are actually locked to the cavity are detuned from the centers of their corresponding cavity modes, and the detuning has opposite sign at longer and shorter wavelengths. The amplitude noise converted by the cavity modes as the frequencies of comb lines jitter around the cavity resonances has opposite phase for comb lines symmetrically red and blue detuned from the locking point. Thus, their noise contribution cancels to a large extent. This is not the case when both locks are engaged, and most of the comb lines contribute equally to the amplitude noise.

For comparison, Fig. 4(b) shows the spectrum transmitted through the cavity around 3900 nm, i.e. away from the maximum of cavity finesse. The FSR changes continuously in this wavelength range, so that only a limited part of the incident comb spectrum can be coupled into the cavity. The blue dashed curves show the spectra with only the pump laser lock enabled, the red dotted and black dash-dotted curves show the spectra at the locking points chosen for pump laser and OPO locks, respectively, and the solid blue curve shows the spectrum when both locks are enabled (the latter multiplied by 1.5). Clearly, a large portion of the spectrum at longer wavelengths (where the FSR changes more rapidly as a function of wavelength) is rejected by the cavity. In order to allow stable transmission the locking points must be spectrally separated by a much smaller gap than in the previous case, namely 30 nm instead of 120 nm. This choice was found optimum to maximize cavity transmission around 3.9 µm for measurements of $N_2O$. The upper panel of Fig. 4(b) shows the detuning of the comb lines from cavity resonances as a function of wavelength.

To demonstrate the advantage of the two-point lock, Fig. 4(c) shows six spectra taken at 8 s intervals with the two-point lock (lower set of curves), only the pump laser lock enabled (middle set of curves, offset for clarity), and only the OPO lock enabled (upper set of curves). It is clear that when only one of the locks is engaged the drift of the spectrum is much larger and the repeatability of the cavity transmission becomes much worse, as only one particular group of comb lines is locked to the cavity.

The spectra shown in Figs 4(a) and 4(b) are used to calculate the detuning of comb modes from the cavity modes as a function of wavelength for a given choice of locking points. The measurement of cavity transmission with a single-point lock reveals the maximum available power at the given wavelength, since the comb lines are locked to the centers of cavity resonances there. With the two-point lock enabled, one can determine the offset between cavity and comb modes from the fractional change of the transmitted power at the given wavelength (by comparing the peak power under a single-point lock to the power at the same wavelength when the two-point lock is enabled). The cavity mode has a Lorentzian line shape with a half width given by $\Gamma_L = FSR/(2F)$. The fractional change of FSR in a given wavelength range is orders of magnitude smaller than the fractional change of the finesse (given by the fit in Fig. 2), so it can be neglected to the first order. Thus one can calculate the frequency detuning of a comb line from the center of the cavity mode at each wavelength using

$$\Delta v = \Gamma_L \sqrt{\frac{P_1 - P_2}{P_2}}, \tag{6}$$

where $P_1$ and $P_2$ are the power at a given wavelength for one- and two-point lock, respectively. The blue circular markers in the upper panels of Figs 4(a) and (b) show this detuning calculated according to Eq. (6) as a function of wavelength for the two locking schemes; the red curves show 3$^{rd}$ order polynomial fits. The value of the detuning, expressed in Hz, is used for calculations of molecular line shapes in Eq. (4). This method does not require *a priori* knowledge of the dispersion properties of the mirrors, but it does not provide information about the sign of the detuning. However, the latter can be easily determined from the sign of the overshoots in molecular absorption lines.

### *3.4. Detection system and signal acquisition*

The resolution of the Fourier transform interferometer is set to 800 MHz, which is sufficient to resolve pressure broadened absorption lines. One interferogram is acquired in 0.5 s, and the subsequent FFT yields a spectrum containing 6000 spectral elements. For measurements of absorption spectra of $C_2H_2$, $CH_4$ and $N_2O$ in nitrogen ($N_2$) the cavity is filled with the gas mixture to the desired pressure, and the analyte spectrum is recorded while constant flow of the gas through the cavity is maintained. After the analyte spectrum is taken, the cavity is evacuated and then refilled to the same pressure with pure $N_2$ and a reference spectrum is recorded. For measurements of $H_2O_2$, first a reference spectrum is measured with cavity filled with $N_2$, and next the glass container is filled with 27% solution of $H_2O_2$ in water. In each case, the analyte spectrum is normalized to the reference spectrum, and a model spectrum is fit to the data. A polynomial baseline and sinusoidal etalon fringes are also fit in order to be removed from the final spectrum. The baseline originates from a slight drift of the laser spectrum due to the instability in the locking servo [as shown in Fig. 4(c)]; small etalon fringes are caused by multiple reflections in the $LiNbO_3$ crystal in the OPO cavity and in the beamsplitter used to pick off the cavity reflected beam.

High absorption sensitivity is obtained by the combination of fast-scanning of the FTS mirror and the autobalancing detection [40]. The noise in cavity transmission has roughly a 1/$f$ dependence, resulting from the low-pass filtering action of the cavity (with a corner frequency equal to the cavity linewidth). Thus the high carrier frequency of the interferogram (200 kHz) places the laser spectrum after FFT in the region of lower noise. The remaining noise is suppressed by the autobalancing detector, which is based on the Hobbs design [78]. In short, the currents from two photodiodes are subtracted and an active feedback loop keeps the difference of the two currents at zero (by adjusting the current from one of the diodes). Thus, even if the power incident on the two detectors is unequal, the feedback ensures efficient subtraction of the common mode noise, while the signal, which is out of phase in the two output beams of the interferometer, is doubled. In this configuration, the autobalancing detector reduces the amplitude noise around 200 kHz down to the detector noise floor, which is a factor of 3 above the shot noise level for optical power of 180 µW on each detector.

**4. Results**

*4.1. Absorption sensitivity*

We determined the absorption sensitivity by measuring the noise in a ratio of two consecutive reference spectra centered around 3750 nm, such as that shown by the blue solid curve in Fig. 4(a). For a normalized spectrum acquired in 1 s with a resolution of 800 MHz (6000 resolved spectral elements), the standard deviation of the noise in the center of the spectrum is $7.1 \times 10^{-4}$. The minimum detectable absorption is calculated as the ratio of the standard deviation of the noise and the effective interaction length inside the cavity, given by $2FL/\pi$ according to Eq. (5). This yields a sensitivity of $5.4 \times 10^{-9}$ cm$^{-1}$ Hz$^{-1/2}$, corresponding to $6.9 \times 10^{-11}$ cm$^{-1}$ Hz$^{-1/2}$ per spectral element.

*4.2. Commercial gas standards*

In order to demonstrate the performance of the system (in terms of precision, signal-to-noise ratio and validity of the line shape model) we measured spectra of acetylene, methane and nitrous oxide from standardized mixtures in nitrogen at a pressure of 760 torr and room temperature (23 ºC). The available gas mixtures were 100 ppm of $C_2H_2$, 10 ppm of $CH_4$, and 1% of $N_2O$, all in $N_2$. For measurements, $C_2H_2$ was diluted with pure $N_2$ at a ratio of 1:10, $CH_4$ was used directly from the bottle, and $N_2O$ was diluted multiple times to an approximate concentration of 200 ppb within a separate gas mixing cylinder. The black curves in Fig. 5 show the measured spectra (normalized to a reference spectrum) of several combination bands of $C_2H_2$ [Fig. 5(a)], the $2\nu_4$ overtone band of $CH_4$ [Fig. 5(b)], and the $2\nu_1$ overtone band of $N_2O$ [Fig. 5(c)]. We calculate model spectra for the pertinent experimental conditions (pressure and temperature) using spectral line data from the HITRAN database [79] and Eq. (4) for the absorption line shape, and fit them to the experimental data with gas concentration as a fitting parameter, together with a polynomial baseline and sinusoidal etalon fringes. The red curves, inverted and offset for clarity, show the fit spectra (without the baseline and etalons, which are divided out), calculated using the experimentally determined values of the detuning of the comb lines from the cavity modes [shown in upper panels of Figs. 4(a) and (b)]. The residuals of the fits are displayed in the bottom panels below each spectrum. For comparison, we show residuals of fits with the detuning set to zero in the middle panel. Cleary, the full model reproduces the experimental spectra much better than the one that neglects the nonzero detuning. In particular, the overshoots in the absorption lines, clearly visible at shorter wavelengths in the spectra of $C_2H_2$ and $CH_4$ and at longer wavelengths in the spectrum of $N_2O$, are real spectroscopic features and are modeled correctly. The structures remaining in the residuals can be attributed partly to incorrect spectral data in the database (e.g., missing absorption lines), and partly to inaccuracies in the experimentally determined parameters such as the detuning of comb lines from the cavity modes and the finesse (the latter having an uncertainty of 5%). The fit to $N_2O$ spectrum returns a concentration value of 868(12) ppb (the error is one standard deviation of concentration obtained from fits to 50 consecutive spectra), which differs from the expected dilution concentration due to a large uncertainty in the mixing cylinder dilution process. The fit concentrations of 8.77(21) ppm and 8.6(3) ppm obtained for $C_2H_2$ and $CH_4$, respectively, are lower than expected. This discrepancy was found to be systematic, i.e., the concentration of each of these two absorbing gases obtained from fits was linear with the expected concentration, but with a slope of less than unity. This discrepancy might have the same causes as the structure in the residual. For example, a fit to the $CH_4$ spectra shown in Fig. 5(b) in the wavelength range between 3720 and 3850 nm, i.e. not including the strongest absorption lines with largest overshoots, returns concentration of 9.34(24) ppb, i.e. closer to the expected value. Another source of discrepancy might be the residual frequency jitter of comb lines with respect to cavity modes. In the case of cw cavity-enhanced spectroscopy with locked laser, whose linewidth is smaller than the cavity mode width, the noise on transmitted intensity is one-sided, mostly negative going (as the laser frequency jitters around the cavity resonance, the transmitted intensity can only drop from the maximum on-resonance value). Thus one can recover the correct absorption line shape by fitting to the maximum values of the signal. This would be the case also for frequency comb spectroscopy with a comb locked to a cavity and a dispersive element and a detector array used for detection. However, the use of a fast-scanning FTS separates the $1/f$ noise from the signal and the signal is proportional to the average power transmitted through the cavity (and not to its maximum). This effect is not taken into account in our model of the absorption line shapes and is likely the source for underestimating the line strength by a few percent.

The presented measurements of $C_2H_2$ and $CH_4$ demonstrate that in the range of maximum cavity finesse, where the FSR changes least, a spectrum with a bandwidth of 200 nm (limited by the bandwidth of the OPO emission) and resolution of 800 MHz can be recorded in 1 s with a SNR of 500. The cavity can also be used at wavelengths more than 100 nm detuned from the range of maximum finesse, as shown by the measurement of $N_2O$. The line shape model taking into account the detuning of the comb lines from the cavity modes allows recovery of useful spectroscopic data also in this range.

*4.3. Hydrogen peroxide*

Figure 6(a) shows (in black) a normalized spectrum measured at room temperature (23 ºC) and pressure of 630 torr with the flask containing the 27% $H_2O_2$ solution in water connected to the cavity. A small flow of nitrogen was maintained through the glass spacer tube in order to reduce the amount of water vapor in the sample. A sum of model spectra of $H_2O_2$ and $H_2O$ is fit to the data, and the two spectra are shown in red and blue, respectively, inverted and offset for clarity. Spectral data for the water lines is taken from the HITRAN database. However, since data for the $\nu_2 + \nu_6$ intercombination band of $H_2O_2$ is not available in this database, we use absolute line intensities at atmospheric pressure obtained by means of conventional FTS [69, 80, 81]. We have recalculated the intensities to our measurement pressure, which is lower due to the higher altitude at which our laboratory is located. However, the change of pressure broadening is not taken into account, which is the main cause of the structure in the residual, shown below the figure. This systematic discrepancy causes an error in the $H_2O_2$ concentration retrieved by the fit, which we estimate to be 20%. In the spectrum shown in Fig. 6(a) the concentration of $H_2O_2$ is found to be 3.6 ppm in 645 ppm of water. At this low water concentration level, the high absorption sensitivity will allow systematic analysis of $H_2O_2$ absorption lines (preferably performed at lower pressures to decrease broadening of absorption features). We estimate the peak absorption at the strongest $H_2O_2$ line at 3761 nm to be $7 \times 10^{-7}$ cm$^{-1}$ for 1 ppm at 630 torr (from Ref. [69]); this indicates that the noise equivalent concentration detection limit of $H_2O_2$ in the absence of water is 8 ppb at 1 s (1 σ).

Figure 6(b) shows (in black) a spectrum recorded when the container with the $H_2O_2$ solution, as well as the glass tube mirror spacer, are heated to 37 ºC and the cavity is open to air at a pressure of 622 torr (i.e., there is no $N_2$ flow through the glass tube). Under these conditions the concentration of water in the cavity increases significantly. Due to the change of refractive index of water in this wavelength range [82, 83], the locking point for the $f_{ceo}$ lock has to be moved to 3760 nm (i.e. closer to the other locking point), and the transmitted bandwidth is limited at wavelengths above 3770 nm (thus causing increased noise and larger overshoots in absorption lines at these wavelengths). The fit spectra of $H_2O_2$ and $H_2O$ are shown in red and blue, respectively, inverted and offset for clarity, and the concentrations returned by the fits are 5 ppm of $H_2O_2$ and 1.2% of water. Clearly, multiline fitting allows recovery of the $H_2O_2$ spectrum even in the presence of a large amount of water.

In order to determine the concentration detection limit of $H_2O_2$ in the presence of water at a percentage level we measured spectra under the same temperature and pressure conditions as were used for Fig. 6(b), but this time with the container filled with deionized water only. A sum of $H_2O$ and $H_2O_2$ model spectra was fit to 100 consecutive water spectra in the wavelength range with the highest SNR, i.e., between 3720 and 3790 nm. The mean water concentration returned by the fits was 2.83(6)%, while the mean $H_2O_2$ concentration was 75 ppb with standard deviation of 130 ppb. Thus, the detection limit of $H_2O_2$ in the presence of almost 3% of water is 130 ppb. The reason why this value is worse than that determined from the background noise level (8 ppb, see above) is the structure in the fit residual after the model water spectrum is subtracted off. Therefore, the detection limit in the presence of water can be improved by refining the line shape model, using more precise spectral data for the water lines, and by reducing the drift of the baseline in the experiment.

## 5. Conclusions

We have demonstrated for the first time cavity-enhanced direct frequency comb spectroscopy in the mid-infrared wavelength region, proving that sensitive real-time optical detection of hydrogen peroxide for breath analysis applications is now possible. A two-point locking scheme is used to stabilize the mid-IR OPO to a cavity with a maximum finesse of 3800 at 3.815 μm. In the proximity of this wavelength the full bandwidth of the OPO (200 nm) is simultaneously transmitted through the cavity. At further detuned wavelengths the change of cavity FSR due to dispersion in the mirror coatings limits the bandwidth of the transmitted spectrum, which is still more than 100 nm wide. We obtain high absorption sensitivity of $5.4 \times 10^{-9}$ cm$^{-1}$ Hz$^{-1/2}$ through the use of a fast-scanning Fourier transform spectrometer with an autobalancing detector, which reduces the amplitude noise caused by the narrow cavity modes down to the detector noise level. A new line shape model taking into account the detuning of the comb lines from the cavity modes is presented and fit to the measured spectra of acetylene, methane, and nitrous oxide, reproducing the experimental data well.

With the present system we obtain a detection limit for hydrogen peroxide of 8 ppb in the absence of water, and of 130 ppb in the presence of nearly 3% of water, both in 1 s. Considering the three orders of magnitude shorter acquisition time, this sensitivity is far better than that of any other technique currently used for breath analysis, and allows detection of $H_2O_2$ at levels sufficient for detection of acute respiratory distress syndrome [65, 66]. Highly sensitive detection of $H_2O_2$ is also of interest in atmospheric chemistry applications [84, 85], where the molecule has a significant role as a stratospheric reservoir for $HO_x$ [86, 87], and is associated with biomass burning [88, 89]. The concentration detection limit of our system is not as high as of some existing systems used for atmospheric research [88, 90]; however, our technique has the potential for application in this field due to its capability to monitor simultaneously numerous other molecular species of interest such as methane, acetylene, nitrous oxide, and formaldehyde, all in a single system that can be potentially compact, low cost, and light weight.

**Acknowledgements**

The authors thank Terry Brown for developing the mid-infrared autobalancing detector. A. F. acknowledges a Swedish Research Council postdoctoral fellowship. A. J. F. acknowledges a National Research Council postdoctoral fellowship. This project is supported by AFOSR, DTRA, NIST, and NSF.

National Research Council postdoctoral fellowship. This project is supported by AFOSR, DTRA, NIST, and NSF.

**Figure captions**

FIG. 1. (a) Theoretical molecular absorption line shapes in cavity transmission calculated according to Eq. (4) when the comb lines are locked to the center of cavity modes (black solid curve), and when the detuning between the comb lines and cavity modes is 10 kHz, i.e., 1/3 of the cavity linewidth (red dashed curve), and 20 kHz, i.e., 2/3 of the cavity linewidth (blue dash-dotted curve).

FIG. 2. (a) Schematic of the experimental setup. The mid-infrared optical parametric oscillator is synchronously pumped by a high-power Yb:fiber femtosecond laser. The mid-IR comb is locked to a high finesse cavity containing the gas sample by the use of two-point locking scheme. The cavity reflected light is redirected with a beam splitter (BS) onto a reflection grating, which disperses the spectrum. Two parts of the spectrum are imaged on two photodetectors (PD1 and PD2) in order to create error signals at two different wavelengths. The feedback is sent to pump laser and OPO cavity lengths, respectively. The light transmitted through the cavity is coupled into a fast-scanning Fourier transform spectrometer (FTS). (b) The two outputs of the interferometer are incident on two photodiodes of the autobalancing photodetector (PD4). A nonplanar ring oscillator (NPRO), whose beam is propagating parallel to the mid-IR beam and detected with PD3, is used for frequency scale calibration. The two interferograms are recorder with a fast data acquisition board (DAQ) and stored on a computer.

FIG. 3. The measured cavity finesse (blue circular markers) and a $3^{rd}$ order polynomial fit (red curve).

FIG. 4. (a) Lower panel shows the spectrum transmitted through the cavity around 3750 nm when the two-point lock is enabled (blue solid curve) and when only one of the locks is enabled (red dotted curve – pump laser lock at 3700 nm, black dash-dotted curve – OPO lock at 3820 nm). The blue dashed curves show spectra transmitted through the cavity as the pump laser locking point is tuned across the OPO spectrum. The upper panel shows the detuning of the comb lines from cavity modes calculated from the measured spectra using Eq. (6) (blue circular markers), together with a 3$^{rd}$ order polynomial fit (red curve). Both panels in (b) show corresponding data for a center OPO wavelength of 3920 nm. (c) The drift of the spectrum in cavity transmission during 50 s when the two-point lock is enabled (bottom set of curves), when only the pump laser lock is enabled (middle set of curves, offset for clarity), and when only the OPO lock is enabled (upper set of curves, offset for clarity).

FIG. 5. Experimental spectra of (a) 10 ppm of acetylene and (b) 10 ppm of methane and (c) 0.8 ppm of nitrous oxide at 760 torr of nitrogen at room temperature (black curves). The red curves, inverted and offset for clarity, show model spectra calculated using spectral data from the HITRAN database and the full line shape model presented in this work, with the experimentally determined detuning of the comb lines from the cavity modes. The residuals of the fits using the full model are displayed in the bottom panels; for comparison residuals of fits when the detuning of all comb lines is set to zero are displayed in the middle panels.

FIG 6. (a) Spectrum of hydrogen peroxide and water, shown in black, together with the fitted model spectra (blue – 645 ppm of $H_2O$, red – 3.6 ppm of $H_2O_2$, inverted and offset for clarity), recorded at room temperature at a pressure of 630 torr with nitrogen flow through the glass spacer tube. Lower panel shows the residual of the fit. (b) Spectrum of hydrogen peroxide and water, shown in black, together with the fitted model spectra (blue – 1.2% of $H_2O$, red – 5 ppm of $H_2O_2$, inverted and offset for clarity), recorded at a temperature of 37 ºC in 622 torr of air. Lower panel shows the residual of the fit.

**Figures**

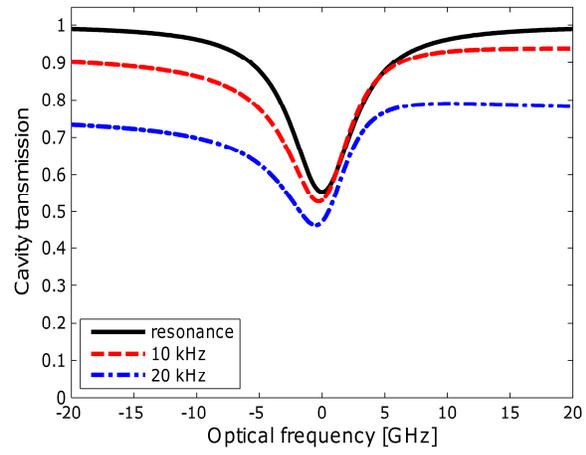

FIG. 1.

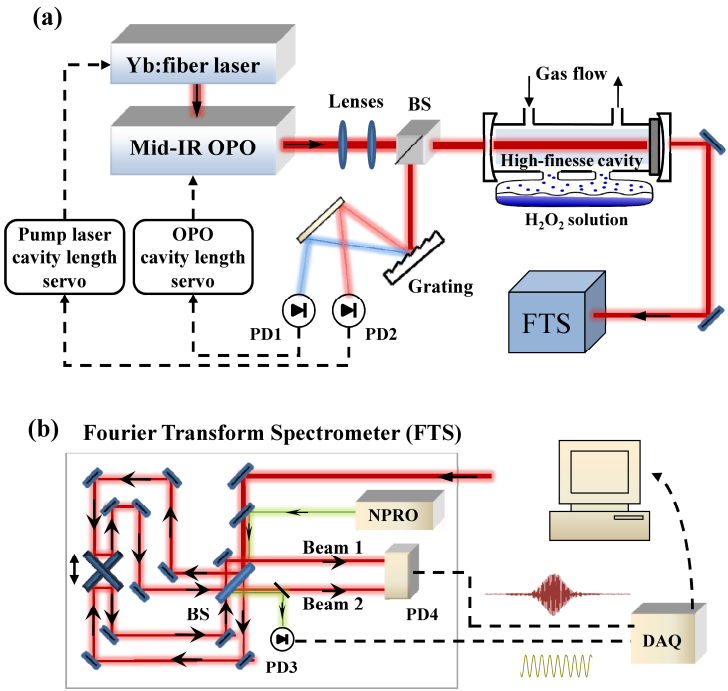

FIG. 2.

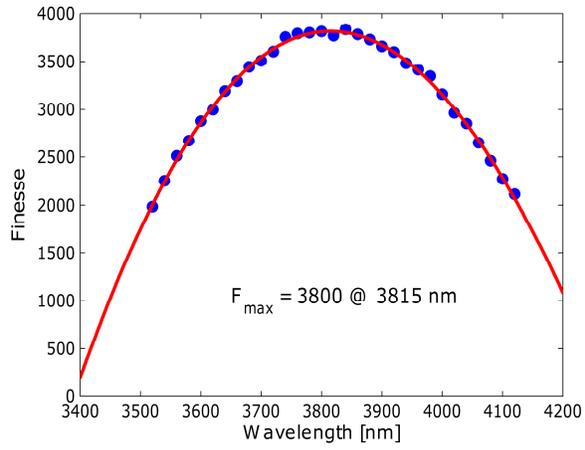

FIG. 3.

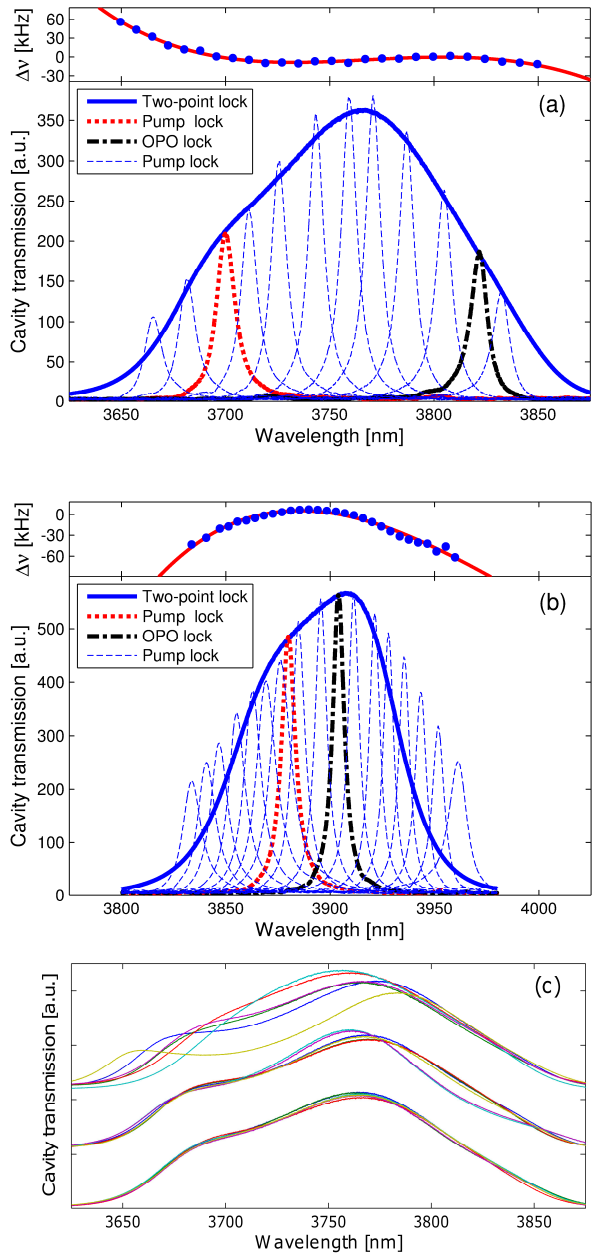

FIG. 4.

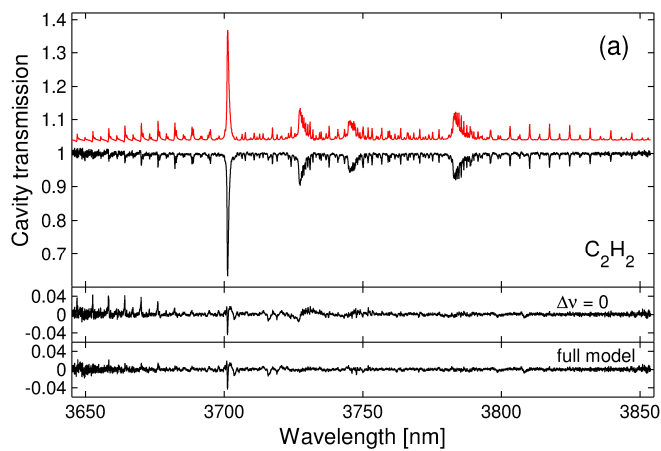

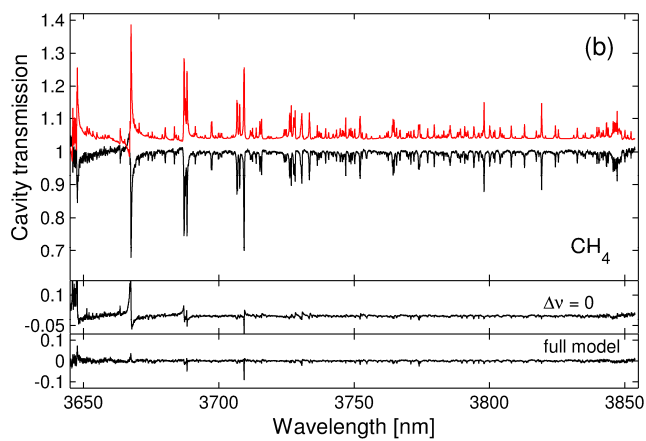

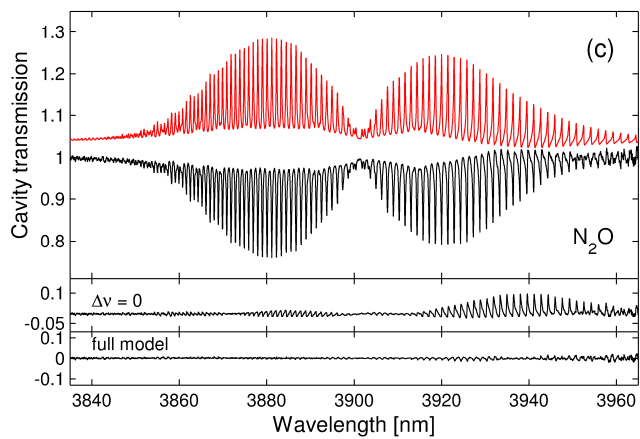

FIG. 5.

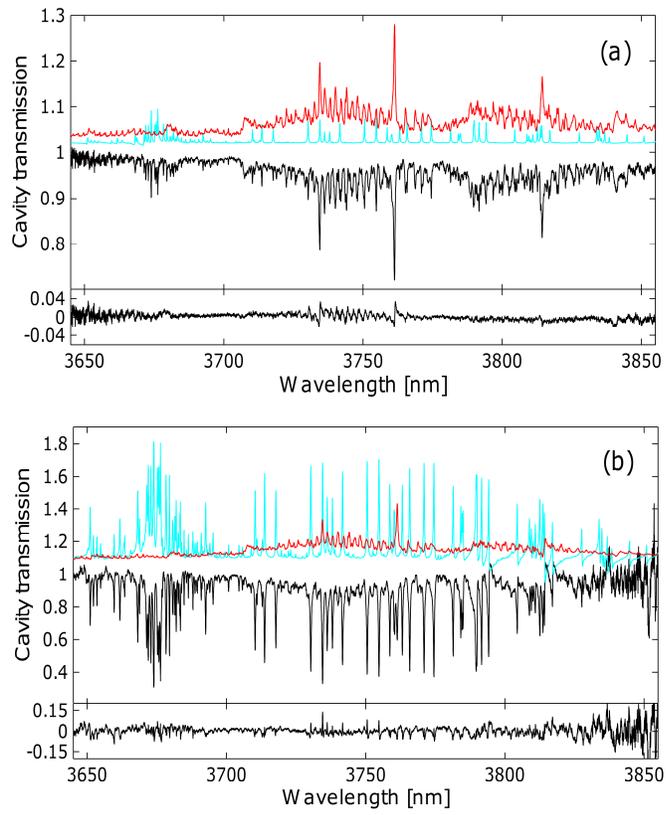

FIG. 6.